\documentclass{iopjournal}

\usepackage{amsmath}
\usepackage{amssymb}
\usepackage[numbers,sort&compress]{natbib}
\usepackage{booktabs}
\usepackage{multirow}
\usepackage{float}
\usepackage{subfig}
\usepackage{enumitem}

\providecommand{\doi}[1]{\href{https://doi.org/#1}{https://doi.org/#1}}

\renewcommand{\articletype}[1]{}

\AtBeginDocument{\rightskip=0pt\relax\leftskip=0pt\relax\parfillskip=0pt plus 1fil\relax}

\begin{document}

\pagestyle{plain}

\articletype{Paper}

\title{Generative Modeling of Complex-Valued Brain MRI Data}

\author{Marco Schlimbach$^1$, Moritz Rempe$^{1,2,3}$, Jessica Mnischek$^1$, Lukas T. Rotkopf$^{2,7}$, Jens Weingarten$^1$, Jens Kleesiek$^{1,2,3,4,5,6}$ and Kevin Kr\"{o}ninger$^1$}

\affil{$^1$Department of Physics, Technical University Dortmund, Otto-Hahn-Stra\ss e 4a, 44227 Dortmund, Germany}

\affil{$^2$Institute for AI in Medicine (IKIM), University Hospital Essen, Girardetstra\ss e 2, 45131 Essen, Germany}

\affil{$^3$Cancer Research Center Cologne Essen (CCCE), University Medicine Essen, Hufelandstra\ss e 55, 45147 Essen, Germany}

\affil{$^4$RACOON Study Group, Site Essen, Essen, Germany}

\affil{$^5$German Cancer Consortium (DKTK), Partner Site Essen, Hufelandstra\ss e 55, 45147 Essen, Germany}

\affil{$^6$Medical Faculty and Faculty of Computer Science, University of Duisburg-Essen, 45141 Essen, Germany}

\affil{$^7$Division of Radiology, German Cancer Research Center (DKFZ), Im Neuenheimer Feld 280, 69120 Heidelberg, Germany}


\begin{abstract}
\textit{Objective.} Standard Magnetic Resonance Imaging (MRI) reconstruction pipelines discard phase information captured during acquisition, despite evidence that it encodes tissue properties relevant to tumor diagnosis. Current machine learning (ML) approaches inherit this limitation by operating exclusively on reconstructed magnitude images. 
The aim of this study is to build a generative framework which is capable of jointly modeling magnitude and phase information of complex-valued MRI scans.
\textit{Approach.} The proposed generative framework combines a conditional variational autoencoder, which compresses complex-valued MRI scans into compact latent representations while preserving phase coherence,
with a flow-matching-based generative model. Synthetic sample quality is assessed via a real-versus-synthetic classifier and by training downstream classifiers on synthetic data for abnormal tissue detection.
\textit{Main results.} The autoencoder preserves phase coherence above 0.997. Real-versus-synthetic classification yields low AUROC values between 0.50 and 0.66 across all acquisition sequences, 
indicating generated samples are difficult to distinguish from real data.
In downstream normal-versus-abnormal classification, classifiers trained entirely on synthetic data achieve an AUROC of 0.880, 
surpassing the real-data baseline of 0.842 on a publicly available dataset (fastMRI). This advantage persists on an independent external test set from a different institution with biopsy-confirmed labels.
\textit{Significance.} The proposed framework demonstrates the feasibility of jointly modeling magnitude and phase information for normal and abnormal complex-valued brain MRI data.
Beyond synthetic data generation, it establishes a foundation for the usage of complete brain MRI information
 in future diagnostic applications and enables systematic investigation of how magnitude and phase jointly encode pathology-specific features.
\end{abstract}

\thispagestyle{empty}

\section{Introduction}
In 2022, more than $320,000$ new cases of brain and central nervous system cancer were reported worldwide \citep{bray2024_globocan}. Precise classification of these tumors is essential, as the diagnosis directly influences clinical decision-making and subsequent treatment. \\
The diagnostic process typically starts with Magnetic Resonance Imaging (MRI), which is the preferred imaging modality in tumor diagnosis due to its ability to provide good soft tissue contrast without exposing patients to ionizing radiation. Although MRI is effective in verifying the existence and location of a tumor, identifying its exact type from imaging alone is often unreliable. Studies confirm this limitation by showing high mismatch rates  between initial MRI-based diagnoses and final histopathological results. Approximately 30\% of MRI diagnoses for brain tumors differ in ways that would affect the treatment plan, with similar misclassification rates of 27.45\% reported for pediatric tumors \citep{ref-pennlund-2022, ref-dixon-2022}. Given these diagnostic discrepancies, histopathological analyses, such as biopsies, remain inevitable in the diagnostic process.
However, brain biopsies carry inherent risks due to their invasive nature, with a systematic review reporting procedure-related mortality of 0.7 to 4\% and overall clinical morbidity between 3 and 13\% \citep{riche2021_systematic}.
Reducing the need for invasive biopsies by accurately determining the tumor type from MRI would thus directly enhance patient safety and treatment outcomes.\\

Many studies have proposed machine learning (ML)
models to improve MRI-based tumor classification \citep{badza2020_cnn, yun2019_radiomics_mlp, mohsen2017_svm, saeedi2023_mri_cnn, abdusalomov2023_mri_dl}, but none of these approaches has yet been sufficient to replace invasive biopsies in clinical practice.
These studies share the limitation that they rely on the final images produced by the MRI scanner, which are the result of a complex reconstruction pipeline and do not preserve the full information originally gathered.
During acquisition, the MRI scanner collects the raw signal of the imaged object in the spatial frequency domain (k-space), 
where each data point is represented as a complex number encoding a specific frequency component of the object.
The k-space data is then transformed into the image domain using an inverse Fourier transform (IFT), 
yielding a complex-valued image whose magnitude and phase now reflect local tissue properties. 
Because MRI is typically performed with multiple receiver coils, 
these coil-specific data are combined into a single final image. Standard combination algorithms mostly rely only on magnitude information,
 as these images are directly interpretable by humans. 
The accompanying phase is typically discarded because for many common acquisition protocols such as short-TR and spin-echo sequences 
it is of limited diagnostic use when inspected visually.
However, this phase information is sensitive to local magnetic susceptibility variations arising from paramagnetic substances such as deoxyhemoglobin and iron, as well as diamagnetic substances such as calcium \citep{haacke2004_swi}. Since brain tumors alter these magnetic properties through intratumoral microhemorrhage and iron deposition \citep{kong2019_itss_glioma, ebrahimpour2025_glioma_swi}, the discarded phase information potentially carries valuable diagnostic information.
Considering phase information can therefore reveal anatomical structures and tumor features that are barely visible or even invisible on magnitude MR images \citep{rauscher2005_phase}. \\

To advance MRI-based tumor diagnosis by including phase information, one promising approach are generative neural networks, since they offer two complementary advantages. 
First, by learning to generate realistic complex-valued MRI samples, such models can serve as a source of synthetic training data. 
This capability is especially relevant since raw MRI datasets are scarce, with pathology-labeled data being particularly rare. 
Second, generative models learn internal feature representations (latent spaces) of their training data, enabling systematic analysis of how magnitude and phase jointly encode pathology-specific features.
Prior work has demonstrated this potential by using structured latent spaces to synthesize and analyze MRI scans of diseased tissue from magnitude images \citep{ref-kleesiek-2021, ref-quiros-2021}. 
However, to date no generative framework for jointly modeling magnitude and phase information exists.
The closest existing work is a diffusion model that synthesizes phase images conditioned on existing magnitude data \citep{rempe2025_phasegen}, 
yet it treats the two components separately rather than modeling them jointly.

In this study, magnitude-only generative approaches are extended by explicitly incorporating phase information into the modeling process.
A generative framework for complex-valued brain MRI data is introduced which models both magnitude and phase information in the image domain for two categories: normal and abnormal brain tissue. Two key challenges are addressed:
\begin{enumerate}
    \item learning a compact latent representation of complex-valued MRI       scans that preserves the coupling between magnitude and phase
          information and provides a structured representation for the subsequent generative model to operate on
    \item conditional generation of normal and abnormal brain tissue under
        label scarcity, addressed through a two-stage training
          strategy which first learns sequence-specific brain anatomy from all
          available data and then finetunes on the smaller labeled subset for
          abnormality-conditioned synthesis.
\end{enumerate}
To evaluate whether the model preserves diagnostically relevant features, downstream classification models are trained on the generated data, with their performance serving as a quantitative measure of synthesis quality. This evaluation is done not only on held-out data from the training distribution but also on an independent external test set with biopsy-confirmed diagnoses from a different institution.
The proposed framework establishes a basis for systematic investigation of complete MRI information through generative approaches.

\section{Materials and Methods}
\begin{figure}[H]
\includegraphics[width=\textwidth]{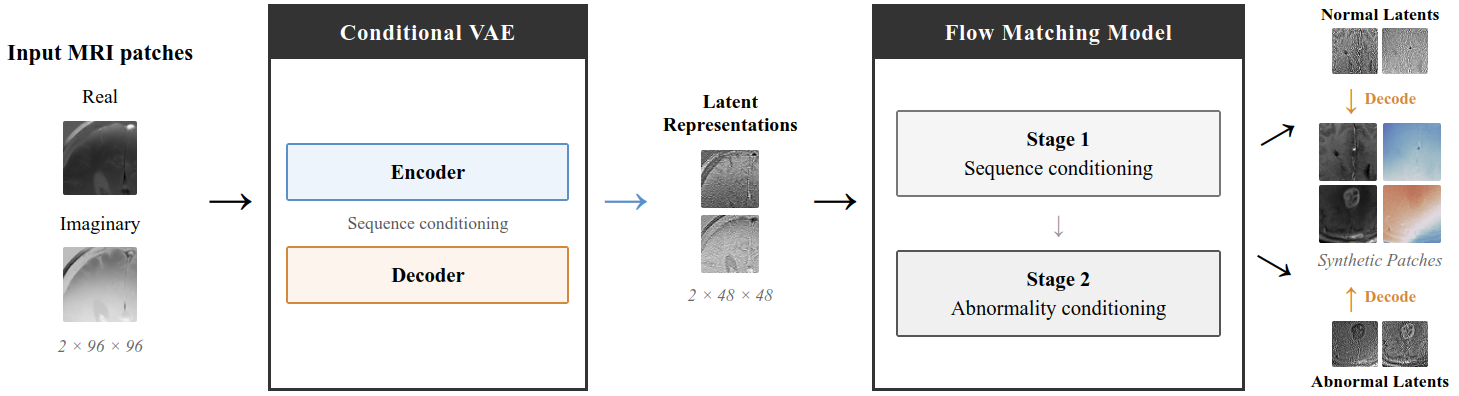}
\caption{Pipeline to generate raw MRI patches. Complex-valued MRI patches $(2 \times 96 \times 96)$ are compressed into latent representations $(2 \times 48 \times 48)$ by a sequence-conditioned conditional variational autoencoder. A flow matching model is then trained in two stages: Stage 1 learns sequence-specific brain anatomy; Stage 2 finetunes on labeled data for normal/abnormal conditioning.\label{fig:pipeline}}
\end{figure}

The workflow for generating complex-valued MRI patches consists of three sequential stages, as depicted in Figure \ref{fig:pipeline}.

\noindent First, an autoencoder is trained to compress complex-valued MRI patches into compact latent representations. These patches are smaller image crops extracted from the full MRI volumes, each represented as a two-channel input where one channel encodes the real part and the other the imaginary part of the complex-valued data. The latent representations then serve as training data for a conditional flow matching model, whose training proceeds in two stages: in the first stage, the model learns to generate sequence-specific brain patches. In the second stage, the pretrained model is adapted to additionally condition on an abnormality label, enabling targeted generation of normal and abnormal tissue. To evaluate the quality of the generated data, classifiers are trained at each stage. After the first stage, a classifier is trained to distinguish real from synthetic latent representations, providing a direct measure of how closely the generated samples match the true data distribution. After the second stage, classifiers are trained on the downstream task of distinguishing normal from abnormal scans using different combinations of real and synthetic data. By comparing the performance of classifiers trained with synthetic data to those trained with real data, the diagnostic utility of the generated samples can be assessed.
Each component is described in detail in the following subsections.
\subsection{Data}
\subsubsection{Source and Labeling}
All models were trained on the fastMRI brain dataset \citep{ref-zbontar-2018}, which comprises multi-coil k-space data of 6,970 brain MRI scans acquired at 1.5T and 3T. 
Five labels are provided by the dataset to specify different acquisition protocols: AXFLAIR (axial FLAIR), AXT1 (axial T1-weighted), AXT1POST (axial post-contrast T1-weighted), AXT1PRE (axial pre-contrast T1-weighted), and AXT2 (axial T2-weighted).
Note that AXT1 and AXT1PRE both correspond to non-contrast T1-weighted acquisitions, but are treated as distinct labels throughout this work to remain consistent with the original labeling.
Corresponding annotations in the form of bounding boxes marking abnormal regions were obtained from the fastMRI+ dataset \citep{ref-zhao-2022}, which covers 1,000 of these scans. These annotations cover a heterogeneous set of imaging findings, including many incidental or nonspecific abnormalities, and are not linked to biopsy-confirmed final diagnoses. Individual samples can contain multiple bounding boxes, each with its own annotation label.
Several annotation categories had only a limited number of samples,
 which motivated consolidating the annotated findings into two labeled classes: \textit{normal} and \textit{abnormal}. 
 The \textit{normal} class comprises samples annotated with the study-level label ``normal for age'' or the image-level label ``normal variant'' 
 (provided no other image-level label was present). The \textit{abnormal} class comprises samples with at least one annotation of ``mass'', 
 ``extra-axial mass'', ``nonspecific lesion'', ``nonspecific white matter lesion'', ``encephalomalacia'', ``lacunar infarct'', or ``likely cysts''.
Annotated samples with labels outside these two class definitions, such as ``possible artifact'' and ``posttreatment change'', were excluded entirely.
The remaining unannotated samples were assigned to a third category: \textit{unlabeled}.

\subsubsection{Coil Combination and Preprocessing}
Each brain volume in the dataset is acquired with multiple receiver coils that capture the signal at different spatial sensitivities, producing a separate k-space measurement per coil.
To combine these into a single complex-valued volume while preserving phase information, the ESPIRiT algorithm \citep{uecker2014_espirit} was adopted. 
This algorithm estimates coil sensitivity maps that describe how each coil responds to signals from different locations. 
These maps are then used to optimally combine the individual measurements, resulting in a single representation of the 3D volume for each sample.
Within each volume, magnitudes were normalized slice-wise to the range $[0, 1]$, while phase values remained in their original range of $[-\pi, +\pi]$.
The data representation was then converted using real and imaginary components instead of magnitude and phase. This choice avoids the discontinuities that arise when phase values jump from $-\pi$ to $+\pi$, thereby simplifying the subsequent learning tasks.

\subsubsection{Patch Extraction}
\label{sec:patching}
\begin{figure}[H]
\includegraphics[width=\textwidth]{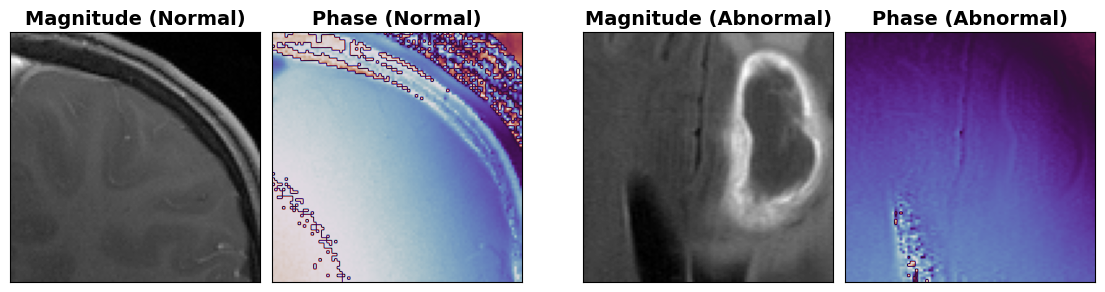}
\caption{Two samples of training data after ESPIRiT combination, IFT and patching for a normal sample (left) and an abnormal sample (right). Magnitude is displayed in grayscale, phase is displayed using a circular colormap spanning $-\pi$ to $+\pi$.\label{fig:train_data}}
\end{figure}

The resulting preprocessed volumes served as the basis for training data preparation. However, using full volumes directly is impractical due to the large number of voxels, which makes training generative models computationally expensive. Furthermore, pathological regions typically occupy only a small fraction of each volume, while the remainder consists of healthy tissue. This imbalance risks shifting the focus of generative models toward healthy anatomy variations rather than the subtle features that characterize pathologies.
To mitigate these issues, 2D patches of size $96 \times 96$ were extracted from the $320 \times 320$ slices instead of processing complete MRI scans. This size ensures that pathological regions occupy a substantial portion of each patch while retaining sufficient surrounding anatomy, thereby emphasizing diagnostically relevant features during training.\\
A magnitude-based brain mask was computed for each slice, and only patches in which at least 80\% of pixels fell within the brain mask were retained. The patch extraction strategy differed between classes: For abnormal volumes, patches were extracted at randomly shifted positions relative to the bounding box annotations, so that the annotated region appears at varying locations within each patch. A patch was extracted if it overlapped by at least 25\% with the bounding box area and the number of patches per bounding box was increased proportionally to the lesion area. Additionally, patches from non-annotated regions within the same volumes were extracted as normal samples. For normal and unlabeled volumes, patches were extracted at random positions across all brain-containing slices. Depending on class and volume, between 10 and 40 patches were extracted per volume. Each complex-valued patch was split into two channels, one for the real part and one for the imaginary part, yielding a final representation of $(2 \times 96 \times 96)$.
The resulting patches were split into train, validation, and test sets at the volume level using a 70/15/15 ratio for the labeled classes and ensuring all patches from a given volume remained in the same set. Since unlabeled samples cannot contribute to the downstream classification evaluation, 95\% of their volumes were allocated to training, with small validation and test portions retained for autoencoder and Stage 1 monitoring.
Figure \ref{fig:train_data} shows examples of the final preprocessed training data for both normal and abnormal cases.
Table \ref{tab:dataset_summary} summarizes the number of patches in each data split.

\unskip

\begin{table}[htbp]
\centering
\caption{MRI Dataset summary by class and sequence. Numbers denote patches.\label{tab:dataset_summary}}
\begin{tabular}{llcccc}
\toprule
\textbf{Class} & \textbf{Sequence} & \textbf{Train} & \textbf{Validation} & \textbf{Test} & \textbf{Total} \\
\midrule
\multirow{4}{*}{Abnormal}
  & AXFLAIR  & 4{,}720 & 1{,}173 & 1{,}503 & 7{,}396 \\
  & AXT1     & 3{,}359 &    376  &    377  & 4{,}112 \\
  & AXT1POST & 2{,}022 &    256  &    475  & 2{,}753 \\
  \cmidrule(l){2-6}
  &          & \textit{10{,}101} & \textit{1{,}805} & \textit{2{,}355} & \textit{14{,}261} \\
\midrule
\multirow{4}{*}{Normal}
  & AXFLAIR  & 4{,}125 & 1{,}020 &   989 & 6{,}134 \\
  & AXT1     & 5{,}820 & 1{,}319 & 1{,}020 & 8{,}159 \\
  & AXT1POST & 3{,}620 &    580  &   900 & 5{,}100 \\
  \cmidrule(l){2-6}
  &          & \textit{13{,}565} & \textit{2{,}919} & \textit{2{,}909} & \textit{19{,}393} \\
\midrule
\multirow{4}{*}{Unlabeled}
  & AXT1POST & 7{,}480 &   192 &   188 & 7{,}860 \\
  & AXT1PRE  & 2{,}350 &    68 &    82 & 2{,}500 \\
  & AXT2     & 25{,}460 &  600 &   720 & 26{,}780 \\
  \cmidrule(l){2-6}
  &          & \textit{35{,}290} & \textit{860} & \textit{990} & \textit{37{,}140} \\
\midrule
\multicolumn{2}{l}{\textbf{Total}} & \textbf{58{,}956} & \textbf{5{,}584} & \textbf{6{,}254} & \textbf{70{,}794} \\
\bottomrule
\end{tabular}
\end{table}

\subsubsection{External Test Data}
\label{sec:external}
To assess whether the downstream classifiers generalize beyond the fastMRI distribution, an unpublished external test set acquired at a separate clinical institution was used. The dataset comprises 96 brain MRI volumes acquired using two pulse sequences, T1-TSE and T2-TIRM,  as summarized in Table~\ref{tab:ext_data}. Unlike the fastMRI+ annotations, each volume carries a biopsy-confirmed label distinguishing healthy tissue from metastases. \\
\begin{table}[htbp]
\centering
\caption{External test set summary by sequence and biopsy-confirmed diagnosis. Numbers denote volumes.\label{tab:ext_data}}
\begin{tabular}{lrrr}
\toprule
\textbf{Sequence} & \textbf{Healthy} & \textbf{Metastasis} & \textbf{Total} \\
\midrule
T1-TSE  &  9 &  7 & 16 \\
T2-TIRM & 43 & 37 & 80 \\
\midrule
\textbf{Total} & \textbf{52} & \textbf{44} & \textbf{96} \\
\bottomrule
\end{tabular}
\end{table}
Several preprocessing steps were applied to spatially align the samples with the fastMRI data. The in-plane dimensions varied across samples and were standardized to $320\times320$ pixels. Dimensions smaller than 320 pixels were zero-filled symmetrically around the k-space center, while dimensions exceeding 320 pixels were symmetrically cropped in the peripheral k-space, effectively applying a low-pass filter that reduces high-frequency noise at the cost of slightly blurring fine detail.
Slices containing facial structures, such as eyes or sinuses, were excluded, as such regions are not present in the fastMRI data.
Because the labels are provided at the volume level, it was not possible to apply the annotation-guided patching strategy described in Section~\ref{sec:patching}. Instead, each slice was divided using a $4\times4$ overlapping grid of ($96\times96$) patches  with a stride of 75 pixels, of which the four center patches were retained to avoid boundary regions dominated by skull or background. Each patch inherits the volume-level label, so that all patches from a healthy volume are labeled as healthy and all patches from a metastasis volume as abnormal. The aggregation from patch-level predictions back to a volume-level score is described in Section \ref{sec:downstream_classification}.

\subsection{Latent Compression}
\label{sec:autoencoder}
The extracted patches from the fastMRI dataset form the input to the first stage of the generative pipeline: an autoencoder that compresses the complex-valued data into compact latent representations. This compression serves two purposes: it reduces the dimensionality that the subsequent flow matching model must handle, and it captures the coupling between the real and imaginary channels. These channels are not independent signals but jointly represent a single complex-valued measurement, where the relationship between them encodes the local phase. Unlike magnitude-only approaches, where a single intensity channel is compressed, this coupling must be explicitly preserved. The architecture, loss function, and evaluation metrics described below are designed accordingly.\\

The architecture is implemented as a ResNet-based~\citep{he2016_resnet} conditional variational autoencoder (CVAE) \citep{sohn2015_cvae}. The encoder maps each $(2 \times 96 \times 96 )$ input patch to a latent representation of size $(2\times 48 \times 48)$, yielding a compression factor of 4. This factor was chosen to retain fine diagnostic detail and the coupling between channels while providing a sufficiently compact input for the flow matching model. The decoder reconstructs the original patch from this compressed encoding.

The loss function comprises three terms
\begin{equation}
\mathcal{L}_{\text{CVAE}} = \mathcal{L}_{\text{recon}} + \lambda_{\text{grad}} \mathcal{L}_{\text{grad}} + \lambda_{\text{KL}} \mathcal{L}_{\text{KL}} ,
\end{equation}
where $\lambda_{\text{grad}}$ and $\lambda_{\text{KL}}$ are weighting factors.
The reconstruction term $\mathcal{L}_{\text{recon}}$ penalizes differences between input and output using the L1 norm, $\mathcal{L}_{\text{grad}}$ preserves edges by matching spatial gradients and the Kullback–Leibler (KL) divergence  term $\mathcal{L}_{\text{KL}}$ regularizes the latent space toward a standard normal distribution to ensure smooth representations suitable for downstream synthesis.\\
\unskip
Since the five MRI acquisition sequences produce distinct global intensity and phase distributions, the autoencoder is conditioned on the acquisition sequence to account for these differences during reconstruction. This is achieved via Feature-wise Linear Modulation (FiLM) conditioning \citep{ref-perez-2018}, which applies learned scale and shift parameters to intermediate activations, allowing the encoder and decoder to adapt to the characteristics of each sequence.
In contrast, conditioning on the class label is not needed at this stage. Since the objective of the autoencoder is purely reconstructive, abnormalities are treated as local anatomical patterns and can be accurately reconstructed without explicit class information. This also allows training a single autoencoder across all sequences, including AXT2 and AXT1PRE, which lack class labels.

Reconstruction quality was assessed using structural similarity index (SSIM) and peak signal-to-noise ratio (PSNR).
To evaluate whether phase information is preserved, the phase coherence between prediction $\hat{x}_i$ and target $x_i$ is measured as
\begin{equation}
\gamma = \frac{|\sum_i \hat{x}_i^* \, x_i|}{\|\hat{x}\| \, \|x\|}
\end{equation}
where $^*$ denotes complex conjugation. The conjugate product at each location yields a vector whose angle reflects the local phase difference between prediction and target. When these phase differences are uniform across the image, the vectors sum constructively and $\gamma$ approaches 1. Since the local magnitude corresponds to the length of each vector, the metric naturally down-weights regions near zero magnitude, where phase is dominated by noise. Furthermore, $\gamma$ is invariant to a constant phase offset applied across all locations, which is a desirable property since global phase offsets in MRI carry no diagnostic information. \\
Once trained, the encoder transforms each patch into a latent representation, which serves as the training data for the flow matching model described in the following section.

\subsection{Flow Matching Model}
\label{section:diffusion}

The latent representations produced by the autoencoder preserve both magnitude and phase information in a compact form. Building on these representations, a flow matching model \citep{lipman2023_flowmatching}  is trained to learn their distribution and to synthesize new samples from it.
Flow matching works by progressively corrupting a clean data sample $x_0$  with Gaussian noise $\epsilon \sim \mathcal{N}(0, I)$ along a linear interpolation path
\begin{equation}
x_t = (1 - t)\,\epsilon + t\,x_0, \quad t \in [0, 1],
\end{equation}
where $t = 0$ corresponds to pure noise and $t = 1$ to the clean sample. A neural network $v_\theta$ is then trained to predict the velocity of this interpolation, which describes the change required to recover a plausible clean sample from a corrupted input $x_t$. The training objective minimizes the squared difference between the networks predicted velocity $v_\theta(x_t, t, c)$ and the true velocity $(x_0 - \epsilon)$:
\begin{equation}
\mathcal{L}_{\text{FM}} = \mathbb{E}_{t,\, x_0,\, \epsilon}\left[\left\| v_\theta(x_t, t, c) - (x_0 - \epsilon) \right\|^2\right],
\end{equation}
where $c$ denotes the conditioning information provided to the network.

Once trained, new samples are generated by starting from pure noise and iteratively applying the predicted changes until a clean latent sample is obtained. \\

 Training proceeds in two stages, each expanding the models capabilities and the conditioning information it receives. In the first stage, the model is conditioned only on the acquisition sequence ($c = c_{\text{seq}}$), which establishes the models ability to generate realistic complex-valued MRI latent representations without any further condition.
In the second stage, the pretrained model is finetuned to additionally condition on the class label ($c = [c_{\text{seq}},\, c_{\text{path}}]$), distinguishing normal from abnormal tissue. This enables targeted generation of each category and provides the basis for investigating what features the model learns to associate with pathologies in the latent space.

\subsubsection{Stage-1: Sequence-Conditioned Training}
The velocity predicting network $v_\theta$ is implemented as a U-Net \citep{ronneberger2015_unet}  with residual blocks and self-attention at the lower spatial resolutions. The noise level and the learned sequence condition are combined into a single conditioning vector, which is injected into each residual block, allowing the network to adapt to the current degree of noise and acquisition sequence. During the first training stage, the sequence condition is randomly dropped for 10\%
of the samples, enabling the model to learn both an unconditional estimate
$v_\theta(x_t, t, \varnothing)$ that captures general brain anatomy and a
sequence-conditioned estimate $v_\theta(x_t, t, c_{\text{seq}})$ that adapts to a specific acquisition sequence.
At sampling, these two estimates are combined via
classifier-free guidance \citep{ref-ho-cfg-2022}
\begin{equation}
\tilde{v} = v_\theta(x_t, t, \varnothing) + w\left(v_\theta(x_t, t, c_{\text{seq}})
- v_\theta(x_t, t, \varnothing)\right),
\end{equation}
where the guidance scale $w \geq 1$ controls the strength of sequence-specific characteristics. Samples are generated by starting from pure noise and iteratively applying the network's predicted velocity using Heun's method until a clean latent sample is obtained. An exponential moving average (EMA) of the model weights is maintained throughout training and used for all inference.
\subsubsection{Stage-2: Abnormality-Conditioned Finetuning}
The first training stage yields a model capable of generating sequence-specific complex-valued MRI latents. To enable targeted generation of normal and abnormal tissue, in Stage 2 the pretrained model is finetuned to additionally condition on the class label that defines if a scan includes an abnormality.
In this stage, all available training data are retained, but class supervision is used only where labels exist (see Table \ref{tab:dataset_summary}). A learnable class embedding is added to the existing conditioning stream, so that each residual block receives a combined vector of noise level, sequence, and abnormality information. Sequences AXT2 and AXT1PRE samples are assigned a dedicated \textit{unknown} token. The \textit{unknown} token allows the model to retain the anatomical and sequence-specific knowledge acquired during Stage~1 for sequences without annotations.
During training, weighted sampling is applied to balance batches across sequences and labels, ensuring that abnormal and normal examples occur at equal frequency despite their imbalanced representation in the dataset.

Directly finetuning all parameters of the model pretrained in Stage 1 risks disturbing the general anatomical representations learned, particularly given the smaller effective set of labeled examples. Training therefore follows a two-phase protocol: in the first phase, all pretrained parameters are frozen and only the newly introduced conditioning parameters are optimized, allowing the model to learn a meaningful embedding. In the second phase, the full model is unfrozen and trained end-to-end, where a lower learning rate is applied to the pretrained parameters  to preserve its learned representations. Classifier-free guidance is used as in Stage 1, but now applied exclusively to the abnormality label while sequence conditioning remains fixed.

\subsection{Evaluation}
The two-stage training approach yields a generative model that can synthesize complex-valued MRI latent representations conditioned on acquisition sequence and abnormality label.
To evaluate these samples quantitatively, classifier-based experiments are conducted at each stage of the flow matching pipeline. After the first stage, a classifier is trained to distinguish real from synthetic latent representations, providing a direct measure of how closely the generated distribution matches the real one. After the second stage, classifiers are trained on the downstream task of separating normal from abnormal tissue. These downstream classifiers are trained on different compositions of real and synthetic data, while evaluation is always performed on a test set consisting exclusively of real scans. This setup reveals whether the synthetic data captures diagnostic features that are relevant for distinguishing normal from abnormal tissue in real clinical cases.

\subsubsection{Latent-Space Evaluation}
\label{sec:real_vs_synth}
If an established classifier is unable to reliably separate real from synthetic samples, this provides evidence that the generated distribution closely matches the real one.
The implemented classifier is ResNet-based and operates directly on the latent representations rather than on decoded images, since the flow matching model is trained to approximate the distribution in this space. Evaluating at this level also avoids artifacts introduced by the decoding step and provides the most direct assessment of generation quality.
Synthetic latent samples are generated using a guidance scale of $w = 1$, with the number of synthetic samples per sequence matched to the corresponding number of real samples. The synthetic samples are split the same way in training, validation and test data. The ResNet classifier is implemented  with squeeze-and-excitation attention and trained using standard regularization techniques including mixup augmentation, cosine learning rate scheduling, and exponential moving average of model weights. Class-balanced sampling is applied to account for the imbalance between real and synthetic samples.
For each acquisition sequence, the classifier is trained five times with different random seeds. To account for variability in the generation process, four independent synthetic test sets are generated for each sequence, each combined with the same fixed real test set. The five trained classifiers are then tested on all four  test sets, and results are reported as mean $\pm$ standard deviation.

\subsubsection{Downstream Classification}
\label{sec:downstream_classification}
While the previous evaluation assesses distributional fidelity, the downstream classification evaluates whether the synthetic samples preserve the diagnostically relevant differences between normal and abnormal tissue that the model learns during Stage 2. To assess this, synthetic latent representations are generated with a guidance scale of $w=2$ and decoded back to the image domain using the autoencoder from Section~\ref{sec:autoencoder}, yielding  synthetic patches of size $(2 \times 96 \times 96)$.
A downstream binary classifier is then trained to distinguish normal from
abnormal patches on this decoded data, so that its performance reflects the
full synthesis pipeline, including any artifacts introduced by the decoding
step.
Two complementary experiments are conducted to evaluate the diagnostic utility of
the synthetic data. The first asks whether synthetic samples can effectively
replace real labeled data. The second investigates whether they can provide additional value when used alongside the full real training set.\\

\textbf{Substitution Experiment.}
The classifier is trained under two conditions using progressively smaller
fractions of the real training set, ranging from 100\% down to 10\% in steps of
10\%. Fractions are defined at the volume level: a real fraction of 70\% retains
70\% of the volumes, along with all patches extracted from those volumes. In the \textit{real-only} condition, the reduced set is used without adding any
additional data, resulting in a smaller overall training set. In the \textit{real + synthetic} condition, the removed real samples
are replaced with class- and sequence-matched synthetic samples so that the total training set size remains constant. The 100\% real-data configuration serves as the baseline. By comparing both conditions across all fractions, this experiment reveals whether synthetic data preserves the diagnostic signal present in the real data. If the synthetic samples capture the relevant distinguishing features between
normal and abnormal tissue, the \textit{real + synthetic} condition should
maintain performance close to the baseline.\\

\textbf{Additive Experiment.}
In this experiment, the full real training set is retained and synthetic data is
progressively added on top, ranging from 10\% to 100\% of the real training set
size in steps of 10\%. This tests whether synthetic samples provide complementary information beyond the real data alone.\\

To isolate the effect of dataset composition from random variation in sample selection, the data fractions are constructed hierarchically. First, a subset containing 10\% of the training data is sampled. Each larger fraction is then formed by adding new samples to the previous subset, so that all samples included at a smaller fraction remain present in every larger fraction. The same procedure is used for the synthetic data.

Since the external test set described in Section~\ref{sec:external} provides labels at the volume level rather than at the patch level, patch-wise predictions must be aggregated to obtain a single score per volume. For each volume, the classifier produces an abnormality probability for every retained center patch, and the volume-level score is computed as the mean of the top 5\% of these probabilities. This emphasizes the most suspicious patches while avoiding excessive averaging over primarily normal tissue.

The downstream classifier shares the same ResNet-based architecture and training procedure used for the real-versus-synthetic evaluation in Section~\ref{sec:real_vs_synth}, adapted to the decoded patch size of
$(2\times 96 \times 96)$. Each configuration is trained 5 times with identical hyperparameters to reduce variance from random initialization and sampling order. An identical augmentation pipeline is applied across all training configurations. The pipeline consists of random horizontal and vertical flips, uniformly sampled 90-degree rotations, random scaling (zoom factor uniformly drawn from $[1.0, 1.2]$ with center-cropping to preserve spatial dimensions), and intensity scaling. Augmenting all conditions heavily helps to isolate the diagnostic content of the synthetic samples from a pure data augmentation effect. The weighted sampler, mixup augmentation, and EMA schedule remain fixed across all runs and conditions.

\subsubsection{Metrics and Model Selection}
Model selection is performed using the validation set described in Table~\ref{tab:dataset_summary}. The selection criterion differs between stages to reflect each stages objective. In Stage 1, the model whose latent representations yielded the \textit{lowest} AUROC for the real-versus-synthetic classifier is selected for fine-tuning, as a lower AUROC indicates that real and synthetic samples are harder to distinguish. In Stage 2, the model achieving the \textit{highest} AUROC on the downstream classification task is selected, reflecting the goal of preserving diagnostically relevant structure in the generated data.
Final performance is evaluated on the held-out test set from Table~\ref{tab:dataset_summary} for both stages, with Stage 2 additionally evaluated on the external test set described in Section~\ref{sec:external}.

\section{Results}
The results are presented following the steps in the generation pipeline.

\subsection{Autoencoder}
\begin{figure}[H]
\includegraphics[width=\textwidth]{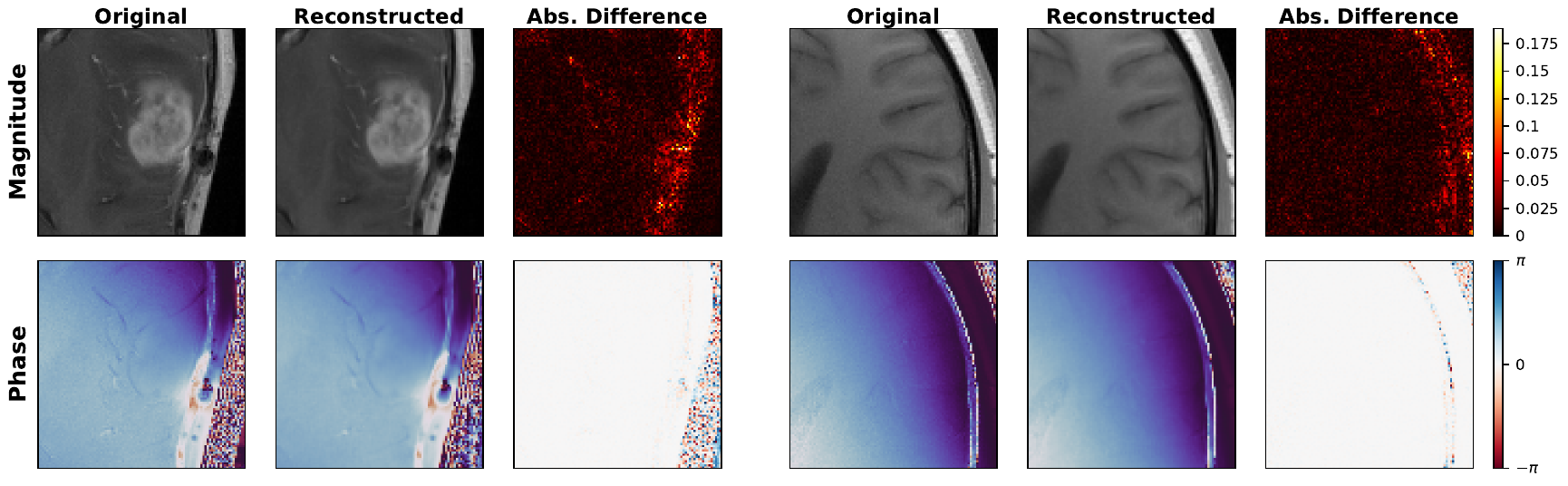}
\caption{Autoencoder reconstruction examples for an abnormal sample (AXT1POST, left) and a normal sample (AXT1PRE, right). Each group shows the original patch, the reconstructed patch, and the absolute difference for both magnitude (top) and phase (bottom).}
\label{fig:AErecon}
\end{figure}

Table~\ref{tab:ae_results} reports phase coherence, magnitude SSIM, and magnitude PSNR for each acquisition sequence. Phase coherence remains above 0.997 across all sequences, indicating that the autoencoder successfully preserves phase information. Magnitude reconstruction quality is high, with SSIM exceeding 0.93 across all sequences and PSNR values ranging from 35.0 to 36.5\,dB, confirming that both structural detail and pixel-level intensity are well retained. \\
\begin{table}[htbp]
\centering
\caption{Per-sequence reconstruction quality on the test set.
Values are reported as mean $\pm$ standard deviation.\label{tab:ae_results}}
\begin{tabular}{lccc}
\toprule
\textbf{Sequence} & \textbf{Phase Coherence ($\gamma$)} & \textbf{Mag.\ SSIM} & \textbf{Mag.\ PSNR (dB)} \\
\midrule
AXFLAIR  & $0.9984 \pm 0.0013$ & $0.955 \pm 0.018$ & $35.8 \pm 2.3$ \\
AXT1     & $0.9983 \pm 0.0008$ & $0.944 \pm 0.029$ & $36.5 \pm 3.3$ \\
AXT1POST & $0.9977 \pm 0.0017$ & $0.932 \pm 0.031$ & $35.0 \pm 2.7$ \\
AXT1PRE  & $0.9986 \pm 0.0009$ & $0.930 \pm 0.035$ & $35.0 \pm 2.9$ \\
AXT2     & $0.9979 \pm 0.0009$ & $0.951 \pm 0.019$ & $35.4 \pm 1.5$ \\
\bottomrule
\end{tabular}
\end{table}
Figure \ref{fig:AErecon} shows example reconstructions, demonstrating that brain tissue intensity is reconstructed with deviations predominantly in the range of $0$–$0.025$ on the $[0, 1]$ normalized magnitude scale. Larger deviations are mostly localized in skull regions, which are diagnostically irrelevant. Phase is reconstructed with near-perfect accuracy, exhibiting deviations close to 0 within the brain. The only exception occurs in regions where the magnitude approaches 0, primarily in the background, where deviations span the full $-\pi$
 to $\pi$ range. This is expected, as phase information in these regions is dominated by noise.
 These results confirm that the compressed latent representations capture both magnitude and phase, providing a reliable basis for the subsequent flow matching model.

\subsection{Flow Matching: Stage 1}

\begin{figure}[H]
\includegraphics[width=\textwidth]{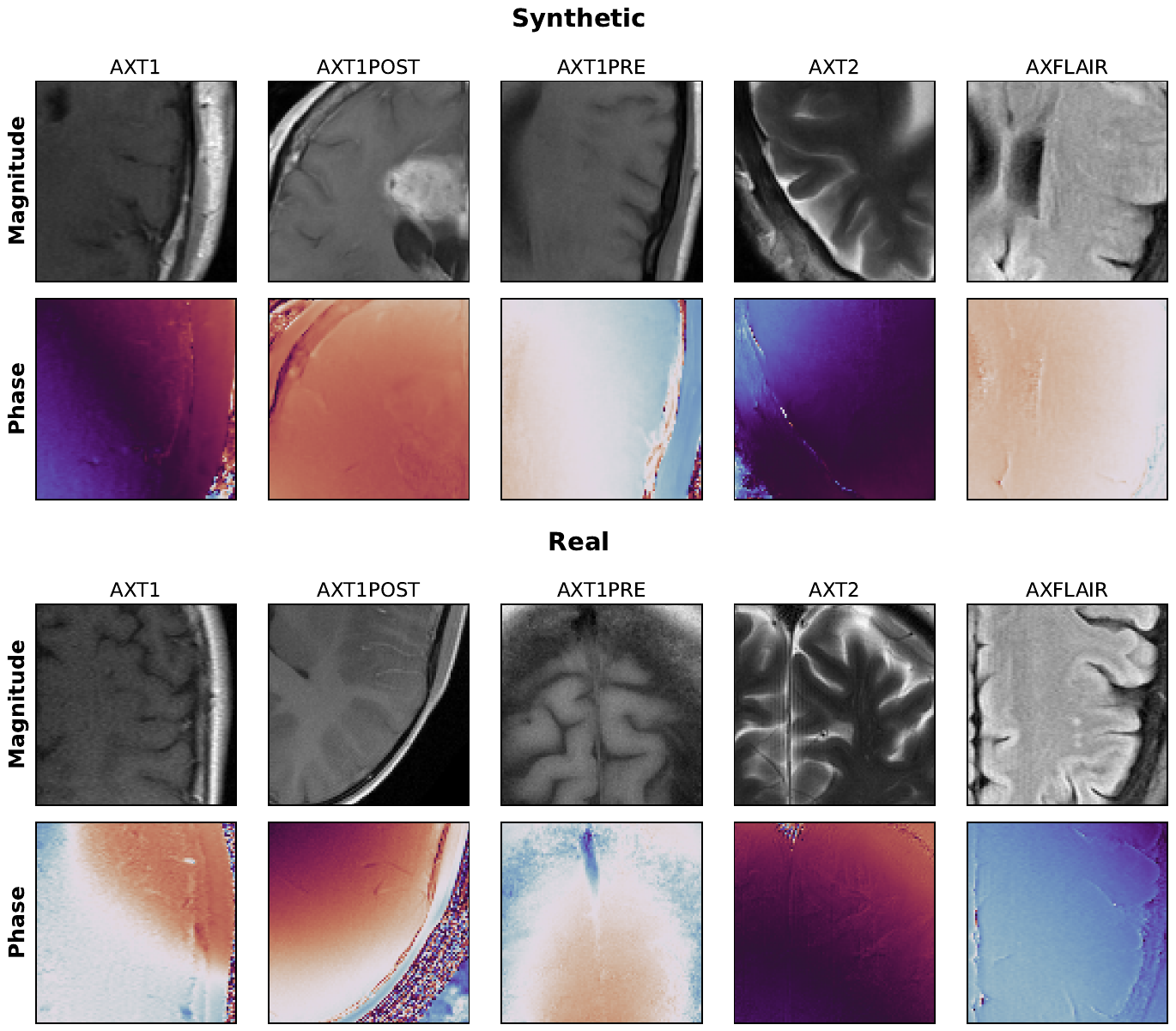}
\caption{Generated and real samples from the Stage 1 flow matching model. Synthetic samples (top) and real samples (bottom) are shown for all five acquisition sequences (AXT1, AXT1POST, AXT1PRE, AXT2, AXFLAIR). For each group, magnitude (top row) and phase (bottom row) are displayed. Magnitude is shown in grayscale, and phase is displayed using a circular colormap spanning from $-\pi$ to $+ \pi$.}
\label{fig:stage1-samples}
\end{figure}

Outputs conditioned on the different sequences can be seen in Figure \ref{fig:stage1-samples}, alongside real samples shown for comparison. The generated samples match the sequence-specific contrast characteristics of the real data and appear anatomically plausible, with consistent magnitude and phase structure.
The results for the latent classification are presented in Table \ref{tab:latent_results}.
Across all sequences, the AUROC values remain close to the chance level of 0.5, indicating that the classifier fails to reliably distinguish real from synthetic latent representations.
Even for the most discriminable sequences AXFLAIR, AXT1, and AXT2, the AUROC reaches only between 0.63 and 0.66.  AXT1POST achieves an AUROC of $0.555 \pm 0.049$, suggesting that the latent distribution of this sequence is captured particularly well. AXT1PRE yields the lowest AUROC at $0.502 \pm 0.014$. However, because this sequence has a considerably smaller sample size, this result should be interpreted with caution.
Overall, the low discriminative ability across all sequences indicates that the generated latent samples closely approximate the real data distribution.

\begin{table}[htbp]
\centering
\caption{AUROC for the real\,vs.\,synthetic latent-space classifier,
reported as mean $\pm$ standard deviation over five classifiers trained
with different random seeds, each evaluated on the four test sets.\label{tab:latent_results}}
\begin{tabular}{lc}
\toprule
\textbf{Sequence} & \textbf{AUROC} \\
\midrule
AXFLAIR  & $0.635 \pm 0.030$ \\
AXT1     & $0.640 \pm 0.026$ \\
AXT1POST & $0.555 \pm 0.049$ \\
AXT1PRE  & $0.502 \pm 0.014$ \\
AXT2     & $0.653 \pm 0.014$ \\
\bottomrule
\end{tabular}
\end{table}

\subsection{Flow Matching Stage 2}
\begin{figure}[H]
\includegraphics[width=\textwidth]{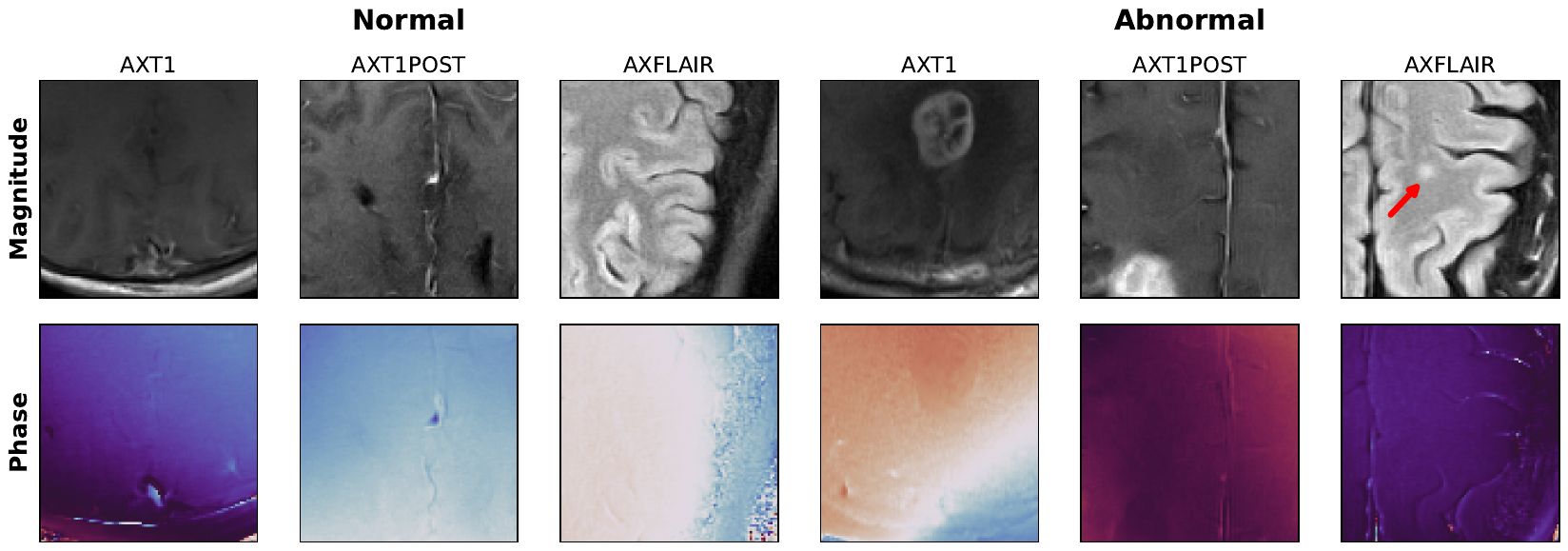}
\caption{Samples generated by the Stage 2 flow matching model, conditioned on normal (left) and abnormal (right) classes for the three labeled acquisition sequences. Magnitude (top) and phase (bottom) are shown. The red arrow highlights a subtle white-matter lesion-like abnormality in the AXFLAIR sample.}
\label{fig:stage2-outputs}
\end{figure}

Examples of outputs conditioned on the normal and abnormal classes are shown in Figure \ref{fig:stage2-outputs}.
In the abnormal samples, tumor- and lesion-like structures are clearly visible and vary in morphology, position, and size. For instance, the abnormal AXT1 sample exhibits a prominent tumor-like mass, whereas the AXFLAIR sample displays a subtle white-matter lesion-like abnormality (highlighted by the red arrow).
The model also produces abnormalities at diverse spatial locations: in the AXT1POST abnormal sample, the pathological region is located at the lower edge and only partially visible within the patch. The corresponding phase images are consistent with their magnitude counterparts and likewise reflect the abnormalities. In both the AXT1 and AXT1POST abnormal phase images, the tumor region manifests as a darker area.
This diversity in abnormality appearance across magnitude and phase channels suggests that the model has learned a rich set of features associated with abnormal tissue.

\subsubsection{Substitution Experiment}
\begin{figure}[H]
\includegraphics[width=\textwidth]{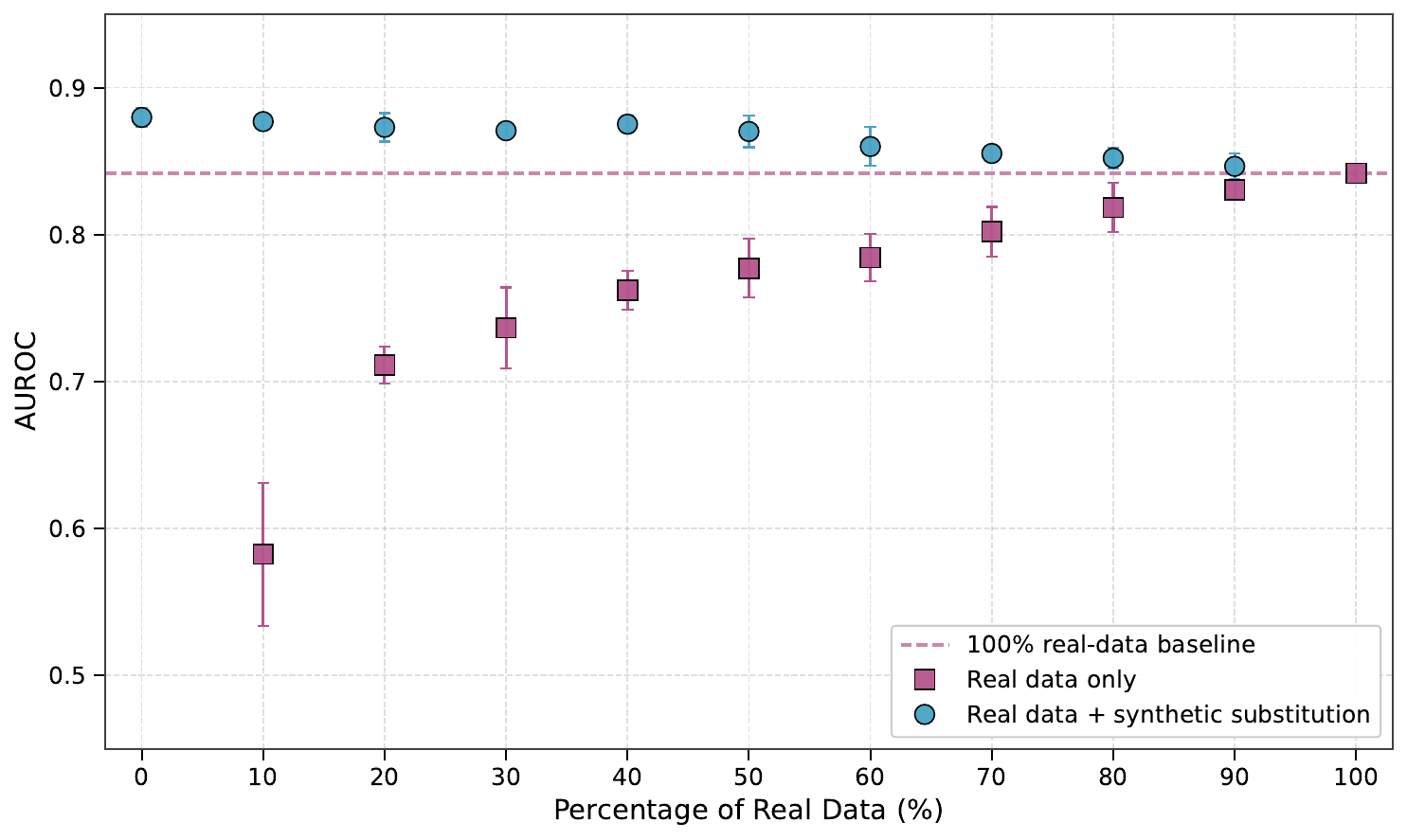}
\caption{Downstream classification AUROC on the fastMRI test set as a function of real data percentage. Values are reported as mean ± standard deviation. The dashed line indicates the 100\% real-data baseline.}
\label{fig:fill_comparison}
\end{figure}

The downstream classification results for the substitution experiment on the fastMRI test set are presented in Figure \ref{fig:fill_comparison}. The 100\% real-data baseline achieves an AUROC of $0.842 \pm 0.006$. Replacing real samples with synthetic data yields AUROC improvements for every composition, not only compared to the corresponding reduced real-data condition, but also relative to the 100\% real-data baseline.
Most notably, the configuration trained entirely on synthetic data (0\% real) achieves the highest AUROC of $0.880 \pm 0.006$ and performance tends to be higher at larger synthetic-data proportions in the training set.

\begin{figure}[H]
\includegraphics[width=\textwidth]{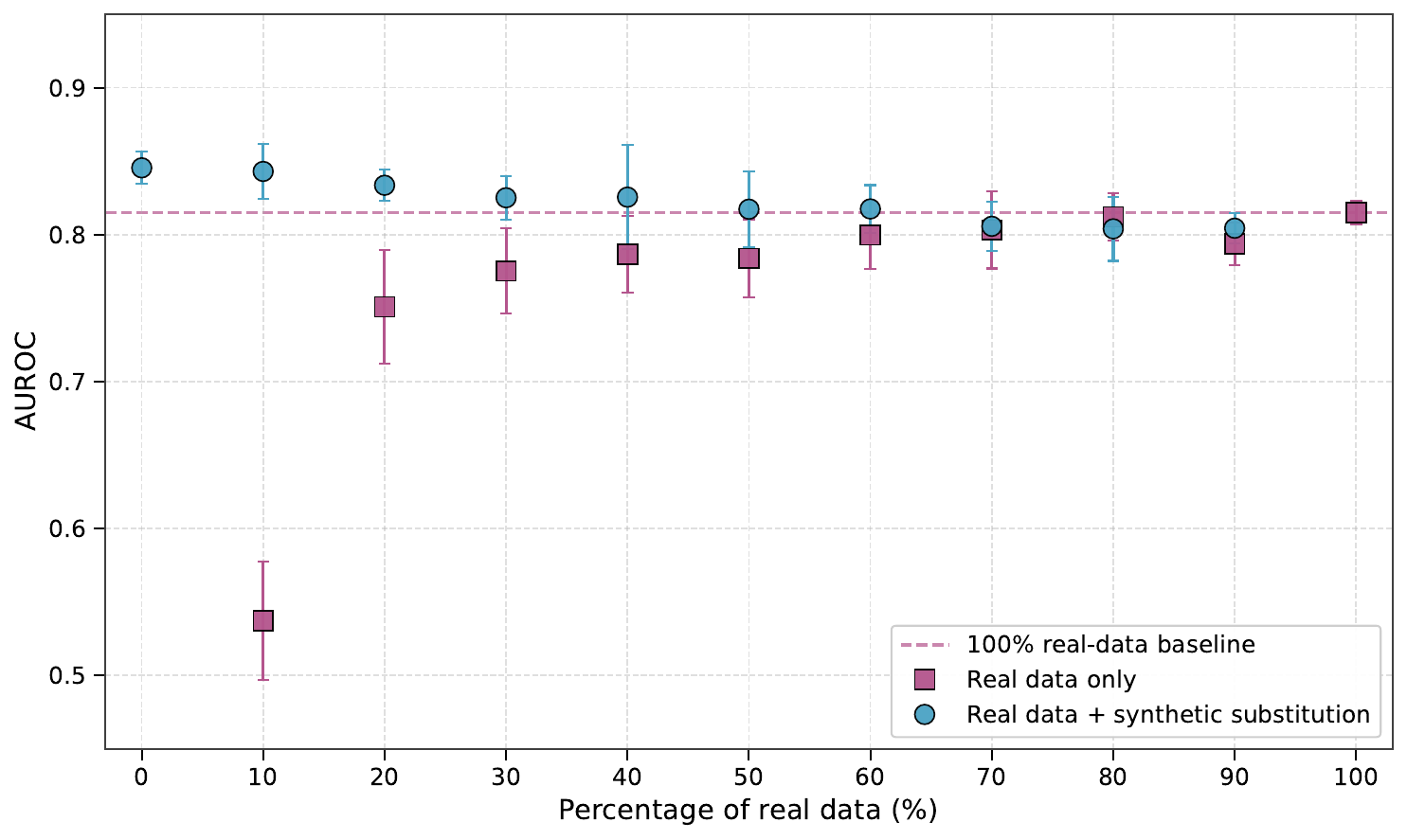}
\caption{Downstream classification AUROC on the external test set as a function of real data percentage. Values are reported as mean ± standard deviation. The dashed line indicates the 100\% real-data baseline.}
\label{fig:fill_ikim}
\end{figure}

The same trend is observed on the external test set (Figure~\ref{fig:fill_ikim}), where the highest AUROC of $0.843 \pm 0.019$ is again achieved by the fully synthetic configuration. Performance decreases as the share of real data increases, approaching the baseline of $0.815 \pm 0.008$ at a composition of 50\% real and 50\% synthetic data. Beyond this point, subsequent compositions fall slightly below the baseline.

\subsubsection{Additive Experiment}

\begin{figure}[H]
    \centering
    \subfloat[\label{fig:plot1}]{%
        \includegraphics[width=0.48\textwidth]{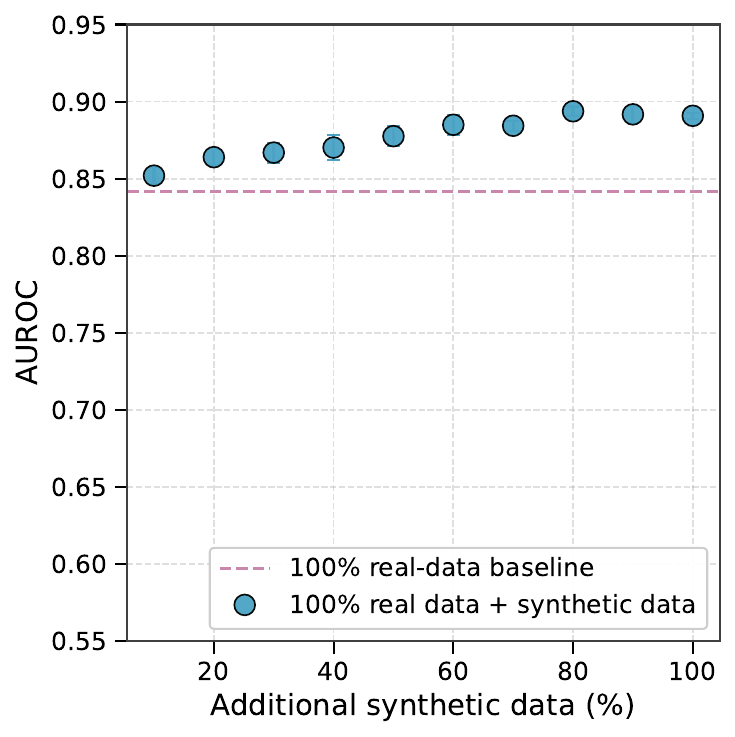}
    }\hfill
    \subfloat[\label{fig:plot2}]{%
        \includegraphics[width=0.48\textwidth]{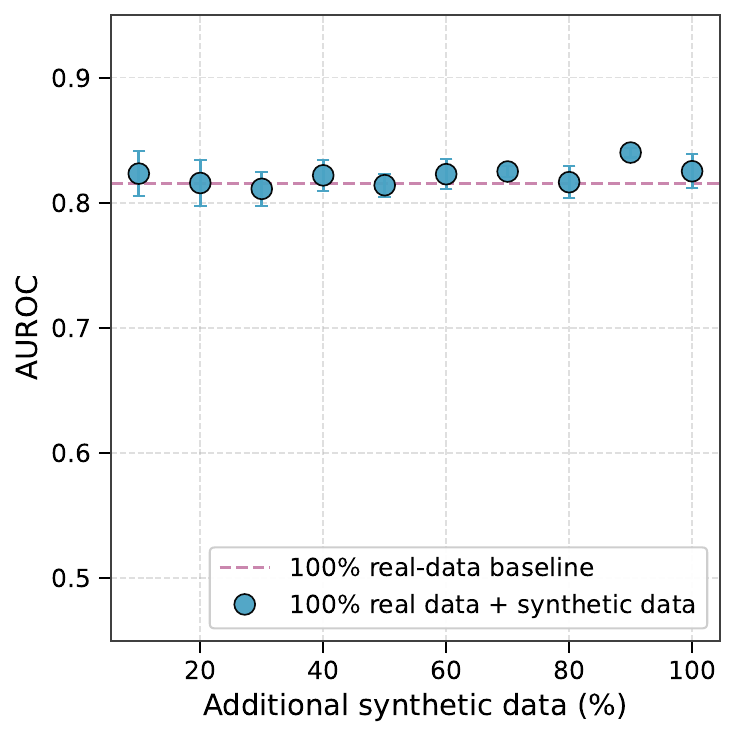}
    }
    \caption{ Downstream classification AUROC when synthetic data is progressively added to the full real training set, evaluated on (a) the fastMRI test set and (b) the external test set. Values are reported as mean ± standard deviation. The dashed line indicates the 100\% real-data baseline.}
    \label{fig:adding}
\end{figure}

The results of the additive experiment, in which synthetic data is progressively added to the full real training set, are shown in Figure \ref{fig:adding}. On the fastMRI test set, performance improves steadily from the baseline of $0.842 \pm 0.006$, reaching $0.894 \pm 0.003$ when 80\% additional synthetic data is included.
At this point the AUROC saturates and further synthetic data does not yield additional gains.
In comparison, this upward trend is not observed for the external test set, where performance stays close to the 100\% real-data baseline of $0.815 \pm 0.008$ with deviations not exceeding $\pm 0.01$ except at the 90\% addition level, which yields an increase to $0.840 \pm 0.004$. Notably, the fully synthetic configuration from the substitution experiment still achieves a higher AUROC than any additive composition on the external test set.

\section{Discussion}

The primary contribution of this work is the generative framework itself. 
To the best of the authors' knowledge, no prior work has jointly modeled magnitude and phase information from complete MRI acquisitions in a generative setting. 
The autoencoder results demonstrate that complex-valued data can be compressed into a latent space while preserving phase coherence above 0.997 across the five tested acquisition sequences,
 confirming that the latent representations accurately encode both magnitude and phase information. The low discriminative ability of the real-versus-synthetic classifier provides direct evidence
  that the flow matching model captures both channels accurately, as systematic errors in either would introduce detectable differences that the classifier could exploit 
  (Table~\ref{tab:latent_results}). Together, these results establish that generative modeling of complex-valued MRI data is feasible and that the proposed pipeline from 
  latent compression to conditional synthesis produces samples that are close to the real data.

Beyond distributional fidelity, the downstream classification experiments show that classifiers trained entirely on synthetic data consistently outperform those trained on real data.
One possible explanation is that the generative model acts as an implicit regularizer \citep{ktena2024_natmed} that attenuates dataset-specific 
characteristics, such as scanner-specific artifacts, acquisition protocol details, and annotation inconsistencies. This interpretation is consistent with related work, where classifiers trained
on synthetic medical images matched or exceeded the performance of those trained exclusively on real data \citep{ktena2024_natmed, khosravi2024_synth}. 
Under this interpretation, replacing real data in the substitution experiment removes these dataset-specific confounders, allowing the classifier to rely on more generalizable abnormality features 
represented by the synthetic data.
This is reflected in the fully synthetic configuration achieving the highest AUROC of $0.880 \pm 0.006$ on the fastMRI test set, surpassing the $0.842 \pm 0.006$ baseline 
(Figure~\ref{fig:fill_comparison}), with performance generally decreasing as the proportion of real data increases. The same pattern holds on the external test set, where the fully
synthetic configuration achieves an AUROC of $0.843 \pm 0.019$ compared to the baseline of $0.815 \pm 0.008$ (Figure~\ref{fig:fill_ikim}). In the additive experiment, however, the full 
real training set remains present, so the classifier retains access to dataset-specific shortcuts. Additional synthetic samples can sharpen the decision boundary within the fastMRI distribution, 
explaining the in-distribution gains up to an AUROC of $0.894 \pm 0.003$ (Figure~\ref{fig:plot1}), but cannot override confounders already learned from the real data, which is consistent with the 
absence of improvement on the external test set (Figure~\ref{fig:plot2}). Notably, the regularization effect seems to extend to the external test set despite a substantial label mismatch: the training
labels group heterogeneous findings without histopathological confirmation, whereas the external data carries biopsy-confirmed metastasis diagnoses. Although metastases may have been present 
in the fastMRI data under the mass or extra-axial mass annotations, no dedicated metastasis signal was ever provided to the model, yet the synthetic-data advantage persists, suggesting that
the retained features generalize across pathology definitions.

However, these results should be interpreted considering two important limitations. First, the external validation is based on a small single-institution dataset of 96 volumes with only one pathology type, 
using sequences that do not directly correspond to those seen during training. Larger multi-site evaluations covering diverse pathologies would be needed to confirm the robustness of these findings. 
Second, the current evaluation is constrained by the available annotations. The broad normal-versus-abnormal distinction does not enable discrimination between specific tumor types, which is the clinically 
relevant task for guiding treatment decisions. Achieving the long-term goal of reducing the need for invasive biopsies will require large datasets with biopsy-confirmed diagnoses and fine-grained
 pathology labels. Such datasets would allow targeted analysis of whether the generative model's latent space encodes tumor-type-specific features, particularly in the combined magnitude and phase information.

Nevertheless, the results demonstrate that generative modeling of complete complex-valued MRI data is feasible and that synthetic samples capture diagnostically meaningful structure,
 establishing a foundation for future work with richer clinical annotations.
\section{Conclusion}
This work introduced a generative framework that, for the first time, jointly models magnitude and phase information from complete complex-valued brain MRI acquisitions. The proposed pipeline, combining a conditional variational autoencoder with a two-stage flow matching model, accurately preserves phase information throughout the entire synthesis process and produces latent representations that are difficult to distinguish from real data.

Downstream classification experiments demonstrate that the generated samples not only retain diagnostically relevant features but yield stronger classification performance than real data alone, an effect that persists on an independent external test set. This suggests that the generative process implicitly regularizes the data by attenuating dataset-specific artifacts while preserving the features most consistently associated with tissue abnormality.

By providing a generative foundation that accurately represents the full information of MRI acquisitions, this work enables systematic investigation of diagnostic features in complete MRI data and offers a framework that can be adapted to other clinical questions and research domains.

\ack{This work was funded by the Bruno \& Helene J\"{o}ster Foundation. The authors declare no conflict of interest.}

\bibliographystyle{unsrtnat}
\bibliography{references}

\begin{thebibliography}{27}
\providecommand{\natexlab}[1]{#1}
\providecommand{\url}[1]{\texttt{#1}}
\expandafter\ifx\csname urlstyle\endcsname\relax
  \providecommand{\doi}[1]{doi: #1}\else
  \providecommand{\doi}{doi: \begingroup \urlstyle{rm}\Url}\fi

\bibitem[Bray et~al.(2024)Bray, Laversanne, Sung, Ferlay, Siegel,
  Soerjomataram, and Jemal]{bray2024_globocan}
Freddie Bray, Mathieu Laversanne, Hyuna Sung, Jacques Ferlay, Rebecca~L.
  Siegel, Isabelle Soerjomataram, and Ahmedin Jemal.
\newblock Global cancer statistics 2022: {GLOBOCAN} estimates of incidence and
  mortality worldwide for 36 cancers in 185 countries.
\newblock \emph{CA: A Cancer Journal for Clinicians}, 74\penalty0 (3):\penalty0
  229--263, 2024.
\newblock \doi{10.3322/caac.21834}.

\bibitem[Pennlund et~al.(2022)Pennlund, Jakola, Skoglund, and
  Ljungqvist]{ref-pennlund-2022}
Anna Pennlund, Asgeir~S. Jakola, Thomas Skoglund, and Johan Ljungqvist.
\newblock A single-centre study of frame-based stereotactic brain biopsies.
\newblock \emph{British Journal of Neurosurgery}, 36\penalty0 (2):\penalty0
  213--216, 2022.
\newblock \doi{10.1080/02688697.2020.1867704}.

\bibitem[Dixon et~al.(2022)Dixon, Jandu, Sidpra, and Mankad]{ref-dixon-2022}
Luke Dixon, Gurpreet~Kaur Jandu, Jai Sidpra, and Kshitij Mankad.
\newblock Diagnostic accuracy of qualitative {MRI} in 550 paediatric brain
  tumours: evaluating current practice in the computational era.
\newblock \emph{Quantitative Imaging in Medicine and Surgery}, 12\penalty0
  (1):\penalty0 131--143, 2022.
\newblock \doi{10.21037/qims-20-1388}.

\bibitem[Riche et~al.(2021)Riche, Amelot, Peyre, Capelle, Carpentier, and
  Mathon]{riche2021_systematic}
Maximilien Riche, Aymeric Amelot, Matthieu Peyre, Laurent Capelle, Alexandre
  Carpentier, and Bertrand Mathon.
\newblock Complications after frame-based stereotactic brain biopsy: a
  systematic review.
\newblock \emph{Neurosurgical Review}, 44\penalty0 (1):\penalty0 301--307,
  2021.
\newblock \doi{10.1007/s10143-019-01234-w}.

\bibitem[Bad{\v{z}}a and Barjaktarovi{\'c}(2020)]{badza2020_cnn}
Milica~M. Bad{\v{z}}a and Marko~{\v{C}}. Barjaktarovi{\'c}.
\newblock Classification of brain tumors from {MRI} images using a
  convolutional neural network.
\newblock \emph{Applied Sciences}, 10\penalty0 (6):\penalty0 1999, 2020.
\newblock \doi{10.3390/app10061999}.

\bibitem[Yun et~al.(2019)Yun, Park, Lee, Ham, Kim, and
  Kim]{yun2019_radiomics_mlp}
Jihye Yun, Ji~Eun Park, Hyunna Lee, Sungwon Ham, Namkug Kim, and Ho~Sung Kim.
\newblock Radiomic features and multilayer perceptron network classifier: A
  robust {MRI} classification strategy for distinguishing glioblastoma from
  primary central nervous system lymphoma.
\newblock \emph{Scientific Reports}, 9:\penalty0 5746, 2019.
\newblock \doi{10.1038/s41598-019-42276-w}.

\bibitem[Mohsen et~al.(2017)Mohsen, El-Dahshan, El-Horbaty, and
  Salem]{mohsen2017_svm}
Heba Mohsen, El-Sayed~A. El-Dahshan, El-Sayed~M. El-Horbaty, and
  Abdel-Badeeh~M. Salem.
\newblock Brain tumor type classification based on support vector machine in
  magnetic resonance images.
\newblock \emph{Annals of the ``Dunarea de Jos'' University of Galati, Fascicle
  II, Mathematics, Physics, Theoretical Mechanics}, 40\penalty0 (1), 2017.

\bibitem[Saeedi et~al.(2023)Saeedi, Rezayi, Keshavarz, and {Niakan
  Kalhori}]{saeedi2023_mri_cnn}
Soheila Saeedi, Sorayya Rezayi, Hamidreza Keshavarz, and Sharareh~R. {Niakan
  Kalhori}.
\newblock {MRI}-based brain tumor detection using convolutional deep learning
  methods and chosen machine learning techniques.
\newblock \emph{BMC Medical Informatics and Decision Making}, 23\penalty0
  (1):\penalty0 16, 2023.
\newblock \doi{10.1186/s12911-023-02114-6}.

\bibitem[Abdusalomov et~al.(2023)Abdusalomov, Mukhiddinov, and
  Whangbo]{abdusalomov2023_mri_dl}
Akmalbek~B. Abdusalomov, Mukhriddin Mukhiddinov, and Taeg~Keun Whangbo.
\newblock Brain tumor detection based on deep learning approaches and magnetic
  resonance imaging.
\newblock \emph{Cancers}, 15\penalty0 (16):\penalty0 4172, 2023.
\newblock \doi{10.3390/cancers15164172}.

\bibitem[Haacke et~al.(2004)Haacke, Xu, Cheng, and Reichenbach]{haacke2004_swi}
E.~Mark Haacke, Yingbiao Xu, Yu-Chung~N. Cheng, and J{\"u}rgen~R. Reichenbach.
\newblock Susceptibility weighted imaging ({SWI}).
\newblock \emph{Magnetic Resonance in Medicine}, 52\penalty0 (3):\penalty0
  612--618, 2004.
\newblock \doi{10.1002/mrm.20198}.

\bibitem[Kong et~al.(2019)Kong, Chen, Zhao, Yao, Fang, Wang, Wang, and
  Li]{kong2019_itss_glioma}
Ling-Wei Kong, Jin Chen, Heng Zhao, Kun Yao, Sheng-Yu Fang, Zheng Wang, Yin-Yan
  Wang, and Shou-Wei Li.
\newblock Intratumoral susceptibility signals reflect biomarker status in
  gliomas.
\newblock \emph{Scientific Reports}, 9\penalty0 (1):\penalty0 17080, 2019.
\newblock \doi{10.1038/s41598-019-53629-w}.

\bibitem[Ebrahimpour et~al.(2025)Ebrahimpour, Ebrahimi, Masoumbeigi, and
  Yeganehdoust]{ebrahimpour2025_glioma_swi}
Anita Ebrahimpour, Tayyebeh Ebrahimi, Mahboubeh Masoumbeigi, and Amin
  Yeganehdoust.
\newblock Magnetic susceptibility-based imaging in gliomas: Insights into tumor
  grading and margin delineation.
\newblock \emph{NMR in Biomedicine}, 38\penalty0 (10):\penalty0 e70140, 2025.
\newblock \doi{10.1002/nbm.70140}.

\bibitem[Rauscher et~al.(2005)Rauscher, Sedlacik, Barth, Mentzel, and
  Reichenbach]{rauscher2005_phase}
Alexander Rauscher, Jan Sedlacik, Markus Barth, Hans-Joachim Mentzel, and
  J{\"u}rgen~R. Reichenbach.
\newblock Magnetic susceptibility-weighted {MR} phase imaging of the human
  brain.
\newblock \emph{AJNR American Journal of Neuroradiology}, 26\penalty0
  (4):\penalty0 736--742, 2005.

\bibitem[Kleesiek et~al.(2021)Kleesiek, Kersjes, Ueltzh{\"o}ffer, Murray,
  Rother, K{\"o}the, and Schlemmer]{ref-kleesiek-2021}
Jens Kleesiek, Benedikt Kersjes, Kai Ueltzh{\"o}ffer, Jacob~M. Murray, Carsten
  Rother, Ullrich K{\"o}the, and Heinz-Peter Schlemmer.
\newblock Discovering digital tumor signatures---using latent code
  representations to manipulate and classify liver lesions.
\newblock \emph{Cancers}, 13\penalty0 (13):\penalty0 3108, 2021.
\newblock \doi{10.3390/cancers13133108}.

\bibitem[Quiros et~al.(2021)Quiros, Murray-Smith, and Yuan]{ref-quiros-2021}
Adalberto~Claudio Quiros, Roderick Murray-Smith, and Ke~Yuan.
\newblock {PathologyGAN}: Learning deep representations of cancer tissue.
\newblock \emph{Machine Learning for Biomedical Imaging}, 1\penalty0 (MIDL 2020
  special issue):\penalty0 1--47, 2021.
\newblock \doi{10.59275/j.melba.2021-gfgg}.

\bibitem[Rempe et~al.(2025)Rempe, H{\"o}rst, Becker, Schlimbach, Rotkopf,
  Kr{\"o}ninger, and Kleesiek]{rempe2025_phasegen}
Moritz Rempe, Fabian H{\"o}rst, Helmut Becker, Marco Schlimbach, Lukas Rotkopf,
  Kevin Kr{\"o}ninger, and Jens Kleesiek.
\newblock {PhaseGen}: A diffusion-based approach for complex-valued {MRI} data
  generation.
\newblock \emph{arXiv preprint arXiv:2504.07560}, 2025.
\newblock \doi{10.48550/arXiv.2504.07560}.

\bibitem[Zbontar et~al.(2018)Zbontar, Knoll, Sriram, Murrell, Huang, Muckley,
  Defazio, Stern, Johnson, Bruno, Parente, Geras, Katsnelson, Chandarana,
  Zhang, Drozdzal, Romero, Rabbat, Vincent, Yakubova, Pinkerton, Wang, Owens,
  Zitnick, Recht, Sodickson, and Lui]{ref-zbontar-2018}
Jure Zbontar, Florian Knoll, Anuroop Sriram, Tullie Murrell, Zhengnan Huang,
  Matthew~J. Muckley, Aaron Defazio, Ruben Stern, Patricia Johnson, Mary Bruno,
  Marc Parente, Krzysztof~J. Geras, Joe Katsnelson, Hersh Chandarana, Zizhao
  Zhang, Michal Drozdzal, Adriana Romero, Michael Rabbat, Pascal Vincent,
  Nafissa Yakubova, James Pinkerton, Duo Wang, Erich Owens, C.~Lawrence
  Zitnick, Michael~P. Recht, Daniel~K. Sodickson, and Yvonne~W. Lui.
\newblock {fastMRI}: An open dataset and benchmarks for accelerated {MRI}.
\newblock \emph{arXiv preprint arXiv:1811.08839}, 2018.
\newblock \doi{10.48550/arXiv.1811.08839}.

\bibitem[Zhao et~al.(2022)Zhao, Yaman, Zhang, Stewart, Dixon, Knoll, Huang,
  Lui, Hansen, and Lungren]{ref-zhao-2022}
Ruiyang Zhao, Burhaneddin Yaman, Yuxin Zhang, Russell Stewart, Austin Dixon,
  Florian Knoll, Zhengnan Huang, Yvonne~W. Lui, Michael~S. Hansen, and
  Matthew~P. Lungren.
\newblock {fastMRI+}, clinical pathology annotations for knee and brain fully
  sampled magnetic resonance imaging data.
\newblock \emph{Scientific Data}, 9\penalty0 (1):\penalty0 152, 2022.
\newblock \doi{10.1038/s41597-022-01255-z}.

\bibitem[Uecker et~al.(2014)Uecker, Lai, Murphy, Virtue, Elad, Pauly,
  Vasanawala, and Lustig]{uecker2014_espirit}
Martin Uecker, Peng Lai, Mark~J. Murphy, Patrick Virtue, Michael Elad, John~M.
  Pauly, Shreyas~S. Vasanawala, and Michael Lustig.
\newblock {ESPIRiT}---an eigenvalue approach to autocalibrating parallel {MRI}:
  Where {SENSE} meets {GRAPPA}.
\newblock \emph{Magnetic Resonance in Medicine}, 71\penalty0 (3):\penalty0
  990--1001, 2014.
\newblock \doi{10.1002/mrm.24751}.

\bibitem[He et~al.(2016)He, Zhang, Ren, and Sun]{he2016_resnet}
Kaiming He, Xiangyu Zhang, Shaoqing Ren, and Jian Sun.
\newblock Deep residual learning for image recognition.
\newblock In \emph{Proceedings of the IEEE Conference on Computer Vision and
  Pattern Recognition (CVPR)}, pages 770--778, 2016.
\newblock \doi{10.1109/CVPR.2016.90}.

\bibitem[Sohn et~al.(2015)Sohn, Lee, and Yan]{sohn2015_cvae}
Kihyuk Sohn, Honglak Lee, and Xinchen Yan.
\newblock Learning structured output representation using deep conditional
  generative models.
\newblock In \emph{Advances in Neural Information Processing Systems
  (NeurIPS)}, pages 3483--3491, 2015.

\bibitem[Perez et~al.(2018)Perez, Strub, de~Vries, Dumoulin, and
  Courville]{ref-perez-2018}
Ethan Perez, Florian Strub, Harm de~Vries, Vincent Dumoulin, and Aaron
  Courville.
\newblock {FiLM}: Visual reasoning with a general conditioning layer.
\newblock In \emph{Proceedings of the AAAI Conference on Artificial
  Intelligence}, volume~32, pages 3942--3951, 2018.
\newblock \doi{10.1609/aaai.v32i1.11671}.

\bibitem[Lipman et~al.(2023)Lipman, Chen, Ben-Hamu, Nickel, and
  Le]{lipman2023_flowmatching}
Yaron Lipman, Ricky T.~Q. Chen, Heli Ben-Hamu, Maximilian Nickel, and Matt Le.
\newblock Flow matching for generative modeling.
\newblock In \emph{Proceedings of the International Conference on Learning
  Representations (ICLR)}, 2023.

\bibitem[Ronneberger et~al.(2015)Ronneberger, Fischer, and
  Brox]{ronneberger2015_unet}
Olaf Ronneberger, Philipp Fischer, and Thomas Brox.
\newblock {U-Net}: Convolutional networks for biomedical image segmentation.
\newblock In \emph{Proceedings of the International Conference on Medical Image
  Computing and Computer-Assisted Intervention (MICCAI)}, pages 234--241, 2015.
\newblock \doi{10.1007/978-3-319-24574-4\textunderscore28}.

\bibitem[Ho and Salimans(2022)]{ref-ho-cfg-2022}
Jonathan Ho and Tim Salimans.
\newblock Classifier-free diffusion guidance.
\newblock \emph{arXiv preprint arXiv:2207.12598}, 2022.
\newblock \doi{10.48550/arXiv.2207.12598}.

\bibitem[Ktena et~al.(2024)Ktena, Wiles, Albuquerque, Rebuffi, Tanno, Roy,
  Azizi, Belgrave, Kohli, Cemgil, Karthikesalingam, and
  Gowal]{ktena2024_natmed}
Ira Ktena, Olivia Wiles, Isabela Albuquerque, Sylvestre-Alvise Rebuffi, Ryutaro
  Tanno, Abhijit~Guha Roy, Shekoofeh Azizi, Danielle Belgrave, Pushmeet Kohli,
  Taylan Cemgil, Alan Karthikesalingam, and Sven Gowal.
\newblock Generative models improve fairness of medical classifiers under
  distribution shifts.
\newblock \emph{Nature Medicine}, 30\penalty0 (4):\penalty0 1166--1173, 2024.
\newblock \doi{10.1038/s41591-024-02838-6}.

\bibitem[Khosravi et~al.(2024)Khosravi, Li, Dapamede, Rouzrokh, Gamble,
  Trivedi, Wyles, Sellergren, Purkayastha, Erickson, and
  Gichoya]{khosravi2024_synth}
Bardia Khosravi, Frank Li, Theo Dapamede, Pouria Rouzrokh, Cooper~U. Gamble,
  Hari~M. Trivedi, Cody~C. Wyles, Andrew~B. Sellergren, Saptarshi Purkayastha,
  Bradley~J. Erickson, and Judy~W. Gichoya.
\newblock Synthetically enhanced: unveiling synthetic data's potential in
  medical imaging research.
\newblock \emph{eBioMedicine}, 104:\penalty0 105174, 2024.
\newblock \doi{10.1016/j.ebiom.2024.105174}.

\end{thebibliography}

\end{document}